\documentclass[aps,prl,showpacs]{revtex4}
\usepackage{graphicx}
\usepackage{amsfonts,amssymb,amsmath}
\usepackage{bm}

\begin{document}

\title{Exact solution of position dependent mass Schr\"{o}dinger
equation by supersymmetric quantum mechanics}
\date{\today}
 
\author{Ramazan Ko\c{c}}
\email{koc@gantep.edu.tr}
\affiliation{Department of Physics, Faculty of Engineering 
University of Gaziantep,  27310 Gaziantep, Turkey}
\author{Hayriye T\"{u}t\"{u}nc\"{u}ler}
\email{tutunculer@gantep.edu.tr}
\affiliation{Department of Physics, Faculty of Engineering 
University of Gaziantep,  27310 Gaziantep, Turkey}

\begin{abstract}
A supersymmetric technique for the solution of the effective mass Schr\"{o}%
dinger equation is proposed. Exact solutions of the Schr\"{o}dinger equation
corresponding to a number of potentials are obtained. The potentials are
fully isospectral with the original potentials. The conditions for the shape
invariance of the potentials are discussed.
\end{abstract}
\keywords{Effective mass, SUSY method}
\pacs{03.65.Ca, 03.65.Ge, 03.65.-w}

\maketitle

\section{Introduction}

The study of Schr\"{o}dinger equation with a position dependent mass has
been the subject of recent interest arising from the study of electronic
properties of semiconductors, quantum dots, liquid crystals and nonuniform
materials in which the carrier effective mass depends on the position \cite%
{serra}. The effective action for a particle with position-dependent mass
has been calculated by back ground method \cite{klein} which in the context
of quantum mechanics is described in the textbook \cite{kleinb}. By the way
the method provides a way to keep manifest the reparametrization invariance
of $\sigma $-models, and is therefore the preferred method for explicit
calculations of the effective actions. Since the position-dependent mass Schr%
\"{o}dinger equation is of considerable importance in condensed matter
physics, we feel that it is necessary to study its solution systematically.

The supersymmetric (SUSY) method is very useful technique for exactly
solvable potentials \cite{cooper}. SUSY has been suggested to give
encouraging results towards an understanding of the degeneracies in atoms
and establish interesting atomic connections. Several authors have
investigated the exact solution of Schr\"{o}dinger equation with
position-dependent mass using various techniques \cite%
{roy,dutra,milan,plas,dekar}.

In this paper we develop the SUSY method to discuss the supersymmetry and
shape invariance of the effective mass potentials. In section 2 we briefly
review the concept of SUSY in quantum mechanics, then we present a modified
SUSY that is applicable to the system with position-dependent mass. After
presenting a general formulation of SUSY to obtain exact solution of the Schr%
\"{o}dinger equation with position-dependent mass, we discuss the condition
of shape invariance of the potentials, in section 3. Working with some
explicit examples like harmonic oscillator, Coulomb and Morse family
potentials we show that, in section 4, such a modification of SUSY is useful
to study effective mass Hamiltonians. The conclusion is given in section 5.

\section{Effective mass Hamiltonian and the SUSY method}

There are several ways to define the kinetic energy operator when the mass
is a function of position. Since momentum and mass operators no longer
commute, the generalization of the Hamiltonian is not trivial. We begin by
defining a general Hermitian effective mass Hamiltonian which is proposed by
von Roos \cite{roos}, 
\begin{equation}
H=\frac{1}{4}\left( m^{\eta }\mathbf{p}m^{\varepsilon }\mathbf{p}m^{\rho
}+m^{\rho }\mathbf{p}m^{\varepsilon }\mathbf{p}m^{\eta }\right) +V(x)
\label{eq:1}
\end{equation}
where $\eta +\varepsilon +\rho = -1$. The limit of the choice of the
parameters $\eta ,\varepsilon $ and $\rho $ is depends on the physical
system. In fact the problem of choice of the parameters has been a long
standing one in quantum mechanics \cite{dutra}. We follow Morrow and
Brownstein \cite{morrow} who have shown that $\eta =\rho $, by comparing
experimental results and/or exact solutions of the some simple models.
Otherwise the wave function is unphysical. Using the restricted Hamiltonian
from the $\eta =\rho $ constraint, we can write%
\begin{equation}
H=\frac{1}{2}\left( m^{-\frac{1}{2}(\varepsilon +1)}\mathbf{p}m^{\varepsilon
}\mathbf{p}m^{-\frac{1}{2}(\varepsilon +1)}\right) +V(x)  \label{eq:2}
\end{equation}%
Let us point out here that the Hamiltonian (\ref{eq:2}) includes the
frequently used form of the Hamiltonians, in the literature \cite{gora},
which can be expressed as 
\begin{subequations}
\begin{eqnarray}
H &=&\frac{1}{2}\left[ \mathbf{p}\frac{1}{m}\mathbf{p}\right] +V(x)\qquad 
\text{for}\quad\varepsilon =-1  \label{eq:3a} \\
H &=&\frac{1}{2}\left[ \frac{1}{\sqrt{m}}p^{2}\frac{1}{\sqrt{m}}\right]
+V(x)\qquad \text{for}\quad\varepsilon =0  \label{eq:3b} \\
H &=&\frac{1}{4}\left[ \frac{1}{m}p^{2}+p^{2}\frac{1}{m}\right] +\frac{%
(\varepsilon -1)^{2}}{8}\frac{m^{\prime 2}}{m^{3}}+\frac{\varepsilon
m^{\prime \prime }}{4m^{2}}+V(x)  \label{eq:3c}
\end{eqnarray}%
Before going further we derive a general effective Hamiltonian for the case
of position-dependent mass. Let us turn our attention to the Hamiltonian(\ref%
{eq:1}). Using the commutation relation 
\end{subequations}
\begin{equation}
\left[ \mathbf{p}, m^{\alpha }\right] =-i\hbar \alpha m^{\alpha -1}
\end{equation}%
one can put the momenta to the right, the Hamiltonian (\ref{eq:1}) takes the
form%
\begin{equation}
H=H_{eff}+V(x).
\end{equation}%
The effective Hamiltonian $H_{eff}$ is given by%
\begin{equation}
H_{eff}=\frac{p^{2}}{2m}-\frac{i\hbar m^{\prime }}{2m^{2}}\mathbf{p}-U(x)
\end{equation}%
where%
\begin{equation}
U(x)=-\frac{\hbar ^{2}}{4m^{3}}\left( 2(\varepsilon +\eta +\varepsilon \eta
+\eta ^{2}+1)m^{\prime 2}-(\varepsilon +1)mm^{\prime \prime }\right) .
\end{equation}%
Note that the effective potential term $U(x)$ can be eliminated by imposing
the constraints over the parameters such that $\varepsilon =-1$ and $\eta =0$%
. In this case the Schr\"{o}dinger equation will not depend on the
parameters.

Our task is now to discuss the solution of the Hamiltonian (\ref{eq:2}) in
the framework of SUSY quantum mechanics. Let us take a look at the SUSY
quantum mechanics for the standard Schr\"{o}dinger equation. The algebra of
SUSY satisfies the following commutation relations: 
\begin{equation}
\left\{ Q^{+},Q^{-}\right\} =H ,\qquad \left[ Q^{\pm },H\right] =0 ,\qquad
\left\{ Q^{\pm },Q^{\pm }\right\} =0  \label{eq:4}
\end{equation}%
The supercharges $Q^{\pm }$ are defined as%
\begin{equation}
Q^{\pm }=B^{\mp }\sigma ^{\pm } ,\qquad B^{\mp }=\frac{1}{\sqrt{2}}\left(
\pm \frac{d}{dx}+\Phi (x)\right)  \label{eq:5}
\end{equation}%
where $\sigma ^{\pm }$ are Pauli matrices and $\Phi (x)$\ is a
superpotential. We may construct a supersymmetric quantum mechanical system
by defining the Hamiltonians such that the relations in (\ref{eq:4}) holds, 
\begin{equation}
H_{\pm }=B^{\mp }B^{\pm }=-\frac{1}{2}\frac{d^{2}}{dx^{2}}+V_{\pm }(x)
\label{eq:6}
\end{equation}%
The partner potentials $V_{\pm }(x)$\ are related to the superpotential $%
\Phi (x)$ by%
\begin{equation}
V_{\pm }(x)=\frac{1}{2}\left( \Phi ^{2}(x)\pm \Phi ^{\prime }(x)\right)
\label{eq:7}
\end{equation}%
The Hamiltonians $H_{+}$ and $H_{-}$ possess the same eigenvalues except for
the zero energy ground state. The zero-energy eigenstate belongs to the $%
H_{-}$, and supersymmetry of quantum system is said to be good SUSY if the
ground state energy of $H_{-}$ (or $H_{+}$) vanishes. In the other case SUSY
is said to be broken. For good SUSY the ground state of $H_{-}$ is given by%
\begin{equation}
\psi _{0}^{-}(x)=C\exp \left( -\int \Phi (x)dx\right)  \label{eq:8}
\end{equation}%
where $C$ is normalization constant. The potentials are shape invariant \cite%
{gen}, that is $V_{+}(x)$ has the same functional form as $V_{-}(x)$ but
different parameters except for an additive constant: 
\begin{equation}
V_{+}(x;a_{0})=R(a_{0})+V_{-}(x;a_{1})  \label{eq:9}
\end{equation}%
where $a_{0}$ and $a_{1}$ stand for the potential parameters in the
supersymmetric partner potentials, and $R(a_{0})$ is a constant. This
property permits an immediate analytical determination of eigenvalues and
eigenfunctions. The eigenvalues and eigenfunctions of the Hamiltonians $%
H_{+} $ and $H_{-}$ are related by 
\begin{subequations}
\begin{eqnarray}
E_{0}^{-}=0 ,\qquad &&E_{1}^{-}=R(a_{0}) ,\qquad E_{n+1}^{-}=E_{n}^{+}
,\qquad E_{n}^{-}=\sum_{k=0}^{n-1}R(a_{k}) ,  \label{eq:10a} \\
&&\psi _{n}^{-}(x;a_{0})=B^{+}(x;a_{0})\psi _{n-1}^{-}(x;a_{1})
\label{eq:10b}
\end{eqnarray}%
In the following we shall modify the standard SUSY technique to the systems
with position-dependent mass. Since the mass is a function of the position,
the supersymmetric operators include mass term. It will be shown that the
following form of the operators are appropriate to study the Hamiltonian (%
\ref{eq:2}), 
\end{subequations}
\begin{subequations}
\begin{eqnarray}
A^{+} &=&\frac{-1}{\sqrt{2}}\left[ m^{-\frac{1}{2}\left( \varepsilon
+1\right) }\frac{d}{dx}m^{\frac{\varepsilon }{2}}\right] +\frac{W(x)}{\sqrt{%
2m}}  \label{eq:11a} \\
A^{-} &=&\frac{1}{\sqrt{2}}\left[ m^{\frac{\varepsilon }{2}}\frac{d}{dx}m^{-%
\frac{1}{2}\left( \varepsilon +1\right) }\right] +\frac{W(x)}{\sqrt{2m}}
\label{eq:11b}
\end{eqnarray}%
where $W(x)$ is the superpotential and $m$ depends on the position. It can
be checked that the supersymmetry relations in (\ref{eq:4}) and (\ref{eq:5})
are satisfied when $B^{\pm }$ are replaced by $A^{\pm }.$ Note that the
operator $\frac{d}{dx}m^{\alpha }$ read as follows: 
\end{subequations}
\begin{equation}
\frac{d}{dx}m^{\alpha }=m^{\alpha }\frac{d}{dx}+\alpha m^{\alpha -1}\frac{dm%
}{dx}
\end{equation}%
We assume that, for good SUSY the ground state wave function belongs to $%
H_{-} $ and is given by 
\begin{equation}
\psi _{0}^{-}(x)=m^{\frac{1}{2}\left( \varepsilon +1\right) }\exp \left(
-\int W(x)dx\right)  \label{eq:12}
\end{equation}%
One can easily check that $A^{-}\psi _{0}^{-}(x)=0.$ The Hamiltonians of
quantum systems with position-dependent mass take the form 
\begin{subequations}
\begin{eqnarray}
H_{-} &=&A^{+}A^{-}=-\frac{1}{2}\left( m^{-\frac{1}{2}\left( \varepsilon
+1\right) }\frac{d}{dx}m^{\varepsilon }\frac{d}{dx}m^{-\frac{1}{2}\left(
\varepsilon +1\right) }\right) +V_{-}(x)  \label{eq:13a} \\
H_{+} &=&A^{-}A^{+}=-\frac{1}{2}\left( m^{\frac{\varepsilon }{2}}\frac{d}{dx}%
m^{-\left( \varepsilon +1\right) }\frac{d}{dx}m^{\frac{\varepsilon }{2}%
}\right) +V_{+}(x)  \label{eq:13b}
\end{eqnarray}%
where the partner potentials are given by 
\end{subequations}
\begin{subequations}
\begin{eqnarray}
V_{-}(x) &=&\frac{1}{2m}\left( W^{2}(x)-W^{\prime }(x)-\frac{\varepsilon
m^{\prime }}{m}W(x)\right)  \label{eq:14a} \\
V_{+}(x) &=&\frac{1}{2m}\left( W^{2}(x)+W^{\prime }(x)-\frac{(\varepsilon
+1)m^{\prime }}{m}W(x)\right)  \notag \\
&&+\frac{(2\varepsilon +1)}{2m}\left( \frac{3}{8}\frac{m^{\prime 2}}{m^{2}}-%
\frac{1}{4}\frac{m^{\prime \prime }}{m}\right)  \label{eq:14b}
\end{eqnarray}%
It is obvious that the kinetic energy terms of the effective mass
Hamiltonian (\ref{eq:2}) and $H_{-}$ are not identical. Therefore the
shape-invariance condition(\ref{eq:4}) does not satisfy for the
position-dependent mass system. If mass is constant it is easy to check that
the physical quantities of the position-dependent mass system reduce to the
physical quantities of the standard system.

In the following section we will discuss the exact solvability of the
position-dependent mass Schr\"{o}dinger equation by using the supersymmetric
procedure given in (\ref{eq:4}) through (\ref{eq:14b}).

\section{Construction of the shape-invariant potentials}

In this section we present a method to construct shape invariant potentials.
It is known that the potentials which satisfy shape invariance condition are
exactly solvable. To illustrate the method we first analyze the construction
of the shape invariant potentials for standard SUSY system. It is well known
that the exactly solvable potentials can be categorized in two groups: the
potentials that their eigenfunctions include Hypergeometric functions and
the confluent hypergeometric functions. The first group consists of the P%
\"{o}schl-Teller, Eckart, Hult\.{e}n potentials etc. and the second group
contains Harmonic oscillator, Coulomb and Morse potentials. The potentials
in each group can be mapped onto each others by point canonical
transformation. This property implies that the superpotential $\Phi (x)$ may
be expressed in two different forms for Natonzon class of potentials by
considering operator transformation applied to the shape-invariant
potentials \cite{cooper}.

The shape invariance condition for constant mass quantum mechanical systems
is given by (\ref{eq:9}). Let us introduce the following general
superpotential: 
\end{subequations}
\begin{equation}
\Phi (x;\lambda _{1};\lambda _{2})=\lambda _{1}r^{\prime }+\lambda _{2}\frac{%
r^{\prime }}{r}+\frac{r^{\prime \prime }}{2r^{\prime }}  \label{eq:15}
\end{equation}%
where $\lambda _{1}$ and $\lambda _{2}$ are arbitrary constants, and $r=r(x)$%
. The harmonic oscillator, Coulomb and Morse type shape invariant potentials
can be constructed by the appropriate choice of $\lambda _{1}$, $\lambda _{2}
$, and $r$. The parameters $\lambda _{1}$ and $\lambda _{2}$ can be
determined from the relation (\ref{eq:9}), while $r$ can be obtained from
the following constraint: 
\begin{equation}
\left( q_{0}+\frac{q_{1}}{r}+\frac{q_{2}}{r^{2}}\right) r^{\prime 2}=1
\label{eq:16}
\end{equation}%
where $q_{1}$, $q_{2}$ and $q_{3}$ are arbitrary constants. One can obtain
the shape-invariant potentials when the parameters of (\ref{eq:16}) are
chosen as: 
\begin{subequations}
\begin{gather}
q_{0}=q_{2}=0,\quad q_{1}=1,\quad r=\frac{1}{4}x^{2},  \notag \\
\lambda _{1}=2\omega ,\qquad \lambda _{2}=-\frac{\ell }{2}-\frac{3}{4}, 
\notag \\
\Phi (x;\ell )=\frac{(\ell +1)}{x}+\omega x,\quad R(\ell )=\ell +\frac{3}{2};
\label{17a} \\
q_{1}=q_{2}=0,\quad q_{0}=1,\quad r=x,  \notag \\
\lambda _{1}=\frac{Ze^{2}}{\ell +1},\qquad \lambda _{2}=(\ell +1),  \notag \\
\Phi (x;\ell )=\frac{(\ell +1)}{x}+\frac{Ze^{2}}{\ell +1},\quad R(\ell )=%
\frac{(Ze^{2})^{2}}{2}\left( \frac{1}{(\ell +1)^{2}}-\frac{1}{\ell ^{2}}%
\right) ;  \label{17b} \\
q_{0}=q_{1}=0,\quad q_{2}=\frac{1}{\alpha ^{2}},\quad r=e^{\alpha x},  \notag
\\
\lambda _{1}=\frac{b}{\alpha },\qquad \lambda _{2}=a,  \notag \\
\Phi (x;a)=\alpha (\frac{1}{2}+\ell )+be^{\alpha x},\quad R(\ell )=\alpha
^{2}(\ell +1).  \label{17c}
\end{gather}%
The superpotentials (\ref{17a}, \ref{17b}, \ref{17c}) are related with the
shape invariant harmonic oscillator, Coulomb and Morse potentials,
respectively. Similar procedure can be followed to obtain shape invariant
family of potentials for the position-dependent mass quantum mechanical
systems.

Let us express the shape-invariance condition for the position-dependent
mass operators $A^{-}$ and $A^{+}$: 
\end{subequations}
\begin{equation}
A^{-}(x,a_{0})A^{+}(x,a_{0})-A^{+}(x,a_{1})A^{-}(x,a_{1})=R(a_{0}).
\label{ed:1}
\end{equation}%
By substituting operators $A^{\pm }$ in (\ref{ed:1}) we obtain the following
shape-invariance condition: 
\begin{eqnarray}
&&\frac{1}{2m}\left( W^{2}(x,a_{0})-W^{2}(x,a_{1})+W^{\prime
}(x,a_{0})+W^{\prime }(x,a_{1})\right) -  \notag \\
&&\frac{m^{\prime }}{2m^{2}}\left( (1+\varepsilon )W(x,a_{0})-\varepsilon
W(x,a_{1})\right) +\frac{(1+2\varepsilon )}{8m^{3}}\left( 3m^{\prime
2}-2mm^{\prime \prime }\right) =R(a_{0})  \label{ed:2}
\end{eqnarray}%
An interesting feature of the SUSY quantum mechanics for the shape-invariant
system satisfying the condition (\ref{ed:2}) is that the entire spectrum can
be determined algebraically. It should be noted that the shape-invariance is
not general integrability condition. In this work we consider the
shape-invariant potentials obtained from the following superpotential: 
\begin{equation}
W(x;\lambda _{1};\lambda _{2})=\lambda _{1}r^{\prime }+\lambda _{2}\frac{%
r^{\prime }}{r}+\frac{r^{\prime \prime }}{2r^{\prime }}+\frac{\varepsilon
m^{\prime }}{2m}  \label{eq:18}
\end{equation}%
with the condition%
\begin{equation}
\left( q_{0}+\frac{q_{1}}{r}+\frac{q_{2}}{r^{2}}\right) r^{\prime 2}=m
\label{eq:19}
\end{equation}%
Using the analogy of standard SUSY method we obtain the following
superpotentials (let $u=\int \sqrt{m}dx$): 
\begin{subequations}
\begin{eqnarray}
&&q_{0}=q_{2}=0 ,\qquad q_{1}=1 ,\qquad r=\frac{1}{4}u^{2} ,  \notag \\
\lambda _{1} &=&2\omega ,\qquad \lambda _{2}=\frac{\ell }{2}+\frac{1}{4} , 
\notag \\
&&W(x;\ell )=\frac{(\ell +1)\sqrt{m}}{u}+\omega \sqrt{m}u+\frac{%
(2\varepsilon +1)m^{\prime }}{4m} ;  \label{eq:20a} \\
&&q_{1}=q_{2}=0 ,\qquad q_{0}=1 ,\qquad r=u ,  \notag \\
\lambda _{1} &=&\frac{Ze^{2}}{\ell +1} ,\qquad \lambda _{2}=-(\ell +1) , 
\notag \\
&&W(x;\ell )=-\frac{(\ell +1)\sqrt{m}}{u}+\frac{Ze^{2}\sqrt{m}}{\ell +1}+%
\frac{(2\varepsilon +1)m^{\prime }}{4m} ;  \label{eq:20b} \\
&&q_{0}=q_{1}=0 ,\qquad q_{2}=\frac{1}{\alpha ^{2}} ,\qquad r=e^{\alpha u} ,
\notag \\
\lambda _{1} &=&\frac{b}{\alpha } ,\qquad \lambda _{2}=\frac{a}{\alpha }-%
\frac{1}{2} ,  \notag \\
&&W(x;a;b)=\left( a+be^{\alpha u}\right) \sqrt{m}+\frac{(2\varepsilon
+1)m^{\prime }}{4m}  \label{eq:20c}
\end{eqnarray}%
Thus we have obtained superpotentials $W(x)$ for the harmonic oscillator,
Coulomb, and Morse family potentials. Note that the square root of the mass $%
m$ should be integrable. In the following section we demonstrate our method
on some explicit examples.

\section{Examples}

In this section we discuss the exact solution of the Schr\"{o}dinger
equation for which the particle mass is given by 
\end{subequations}
\begin{equation}
m=\left( \frac{\delta +x^{2}}{1+x^{2}}\right) ^{2}  \label{eq:21}
\end{equation}%
As we mentioned before the the spectrum of the position-dependent mass
systems and constant mass systems are identical. For the given mass the
function $u$ takes the form%
\begin{equation}
u=\int \sqrt{m}dx=x+(\delta -1)\arctan x  \label{eq:22}
\end{equation}

\subsection{Harmonic oscillator family potential}

The superpotential for the harmonic oscillator potential is given in (\ref%
{eq:20a}). Substitution of (\ref{eq:20a}) into equations (\ref{eq:14a}) and (%
\ref{eq:14b}) leads to 
\begin{eqnarray}
V_{-}(x;\ell ) &=&\frac{\ell (\ell +1)}{2u^{2}}+\frac{1}{2}\omega u^{2}-%
\frac{1}{2}(2\ell +3)\omega +V_{m}  \notag \\
V_{+}(x;\ell ) &=&\frac{\ell (\ell +3)+2}{2u^{2}}+\frac{1}{2}\omega u^{2}-%
\frac{1}{2}(2\ell +1)\omega +V_{m}  \label{eq:23}
\end{eqnarray}%
where $V_{m}$ is given by%
\begin{equation}
V_{m}=\frac{(\varepsilon (2-\varepsilon )+5/4)m^{\prime 2}}{8m^{3}}-\frac{%
(\varepsilon +1/2)m^{\prime \prime }}{4m^{2}}  \label{eq:24}
\end{equation}%
For the mass given in (\ref{eq:21}), $V_{m}$ takes the form%
\begin{equation}
V_{m}=\frac{(\delta -1)(2\varepsilon +1)\left[ -\delta +(2-2\varepsilon
+2\varepsilon \delta )x^{2}+3x^{4}\right] }{2(\delta +x^{2})^{4}}
\label{eq:25}
\end{equation}%
The spectrum of the mass dependent harmonic oscillator family potentials is
the same as the standard harmonic oscillator potential, and it is given by%
\begin{equation}
E=2n\omega  \label{eq:26}
\end{equation}%
Therefore one may obtain isospectral potentials with different masses. The
reason for this is that the SUSY quantum mechanical procedure given here
affects only the potential and leaves the position-dependent mass unchanged.
In the supersymmetric formulation the oscillator potential is singular with
singularity at $x=0$.

\subsection{Coulomb family potentials}

Similarly one can obtain Coulomb family potential by substituting the
superpotential (\ref{eq:20b}) into equations (\ref{eq:14a}) and (\ref{eq:14b}%
): 
\begin{eqnarray}
V_{-}(x;\ell ) &=&\frac{\ell (\ell +1)}{2u^{2}}-\frac{Ze^{2}}{u}+\frac{%
Z^{2}e^{4}}{2(\ell +1)^{2}}+V_{m}  \notag \\
V_{+}(x;\ell ) &=&\frac{(\ell +1)(\ell +2)}{2u^{2}}-\frac{Ze^{2}}{u}+\frac{%
Z^{2}e^{4}}{2(\ell +1)^{2}}+V_{m}  \label{eq:27}
\end{eqnarray}%
The functions $u$ and $V_{m}$ are the same as defined in (\ref{eq:22}) and (%
\ref{eq:24}). Using the standard procedure one can obtain the following
eigenvalues: 
\begin{equation}
E=\frac{Z^{2}e^{4}}{2(\ell +1)^{2}}-\frac{Z^{2}e^{4}}{2(\ell +n+2)^{2}}
\label{eq:28}
\end{equation}%
which is the same as the constant mass Coulomb potential.

\subsection{Morse family potentials}

The final example is the construction of the Morse family potential. Using
the standard procedure we can obtain the following potentials: 
\begin{eqnarray}
V_{-}(x;a) &=&\frac{b}{2}\left( 2a-\alpha \right) e^{\alpha u}+\frac{b^{2}}{2%
}e^{2\alpha u}+\frac{a^{2}}{2}+V_{m}  \notag \\
V_{+}(x;a) &=&\frac{b}{2}\left( 2a+\alpha \right) e^{\alpha u}+\frac{b^{2}}{2%
}e^{2\alpha u}+\frac{a^{2}}{2}+V_{m}  \label{eq:29}
\end{eqnarray}%
with the eigenvalues%
\begin{equation}
E=\frac{1}{2}\left( a^{2}-(a+n\alpha )^{2}\right)  \label{eq:30}
\end{equation}%
The potential exhibits the spectrum of the Morse oscillator potential.

It also can be shown that when $\delta =1$ the potentials (\ref{eq:23}, \ref%
{eq:27}, \ref{eq:29}) reduce to the standard harmonic oscillator, Coulomb,
and Morse potentials, respectively.

\section{Conclusion}

In this paper we have discussed the exact solution of the position-dependent
mass Schr\"{o}dinger equation by using SUSY quantum mechanical method. We
have shown that both, Schr\"{o}dinger equations with different masses and
potentials can exactly be isospectral. Isospectral potentials have identical
spectra, and are ``self isospectral'' in the sense that the potentials have
identical shape \cite{dunne}. These potentials have been constructed by
linear transformation of the ladder operators, in the constant mass SUSY
system \cite{cao}. Detailed properties of isospectral potentials have been
discussed in \cite{cooper1}.

Finally we would like to mention that the method discussed here can be
generalized for the P\"{o}schl-Teller, Eckart, etc. potentials. The systems
with position dependent mass are relevant in many areas of physics, such as
nuclear physics, hetero- structures, and inhomogeneous crystals, that is
quickly developing. It is hoped that the formalism developed here can be
used for the treatment of the quantum mechanical problems with
position-dependent masses.

\end{document}